\documentclass[a4paper,12pt]{article}
\usepackage{amsmath}
\usepackage{amssymb}
\usepackage{setspace}
\usepackage{cite}
\begin{document}

\section*{The Killing vectors and symmetry in $R$-spacetime}
\begin{center}
{T.~Angsachon$^\dag$, R.~Dhanawittayapol$^\ddag$, P.~Cheewaphutthisakun$^\ddag$, S.~N.~Manida$^\sharp$.}
\end{center}
\begin{center}
{\it Department of Physics, Faculty of Science and Technology,\\ Thammasat University, Thailand$^\dag$,}
\end{center}
\begin{center}
{\it Department of Physics, Faculty of Science, Chulalongkorn University, Thailand$^\ddag$,}
\end{center}
\begin{center}
{\it Department of High Energy Physics, Faculty of Physics,\\ Saint-Petersburg State University, Russia$^\sharp$}
\end{center}
\begin{abstract}
In this paper the Killing vector will be constructed for the $R$-spacetime metric. The symmetry transformations corresponding to this vectors are obtained explicitly.
Their coincidence with the transformations of the Poincaré group in a small neighborhood of the world point is shown.
\end{abstract}

{\bf Keywords:} Killing vectors, R-spacetime, symmetry transformation.
\newpage
\section{Introduction}
It is known that \cite{00},\cite{01},  like Minkowski space-time, the space-time of constant curvature  (positive or negative)  has the maximal symmetry.
These spacetime correspondingly are de Sitter or anti-de Sitter spacetime. In the case of anti-de Sitter spacetime metric in Beltrami coordinates there is a limit $c\rightarrow\infty$
which reaches to the metric of $R$-spacetime \cite{02},\cite{03},\cite{04},\cite{05}. In this work we consider this metric and show that the $R$-spacetime is also a spacetime of maximal symmetry.

\section{Killing vectors and generators of symmetry in $R$-spacetime}

The line element of the $R$-spacetime metric can be represented as \cite{03},\cite{04},\cite{05}
\begin{equation}\label{1}
ds^2 = \frac{R^4}{c^4t^4}\left(1-\frac{\delta_{ij}x^ix^j}{R^2}\right)dx_0^2+\frac{2R^2}{c^3t^3}\delta_{ij}x^i dx^j dx^0-\frac{R^2}{c^2t^2}\delta_{ij}dx^i dx^j,
\end{equation}
where indices $i,j$ run over 1,2,3 and carry out summation by repeated indices.

Obviously, the components of this metric tensor are
\begin{equation}\label{2}
g_{00} = \frac{R^4}{c^4t^4}\left(1-\frac{\delta_{ij}x^ix^j}{R^2}\right), \qquad g_{oi} = \frac{R^2}{c^3t^3}x_i, \qquad g_{ij} =  -\frac{R^2}{c^2t^2}\delta_{ij},
\end{equation}
and their inverse elements are
\begin{equation}\label{3}
g^{00} = \frac{c^4t^4}{R^4}, \qquad g^{0i} = \frac{c^3t^3x^i}{R^4}, \qquad g^{ij} = \frac{c^2t^2}{R^2}\left(\frac{x^ix^j}{R^2}-\delta^{ij}\right).
\end{equation}
It is straightforward to show that the non-zero components of Christoffel symbols are
\begin{equation}\label{4}
\Gamma^0_{00} = -\frac{2}{ct}, \qquad \Gamma^i_{0j} = -\frac{\delta^i_j}{ct},
\end{equation}
which is necessary for calculating the Killing vectors.

Let  $\xi_\mu \equiv (\xi_0,\xi_i)$ be the Killing vector satisfying the Killing equations in curved spacetime \cite{06},\cite{07}
\begin{equation}\label{5}
\partial_\mu\xi_\nu+\partial_\nu\xi_\mu = 2\Gamma^\rho_{\mu\nu}\xi_\rho.
\end{equation}
For the case $\mu = \nu = 0$ we can write the Killing equation as
\begin{equation}\label{6}
\partial_0\xi_0 = -\frac{2}{ct}\xi_0.
\end{equation}
The general solution of (\ref{6}) could be expressed as
\begin{equation}\label{7}
\xi_0 = \frac{f(x)}{t^2},
\end{equation}
where $f(x)$ is an arbitrary function of coordinates.

Now let consider the case $\mu = 0, \nu = i$ $:$
\begin{equation}\label{8}
\partial_0\xi_i+\partial_i\xi_0 = -\frac{2}{ct}\xi_i.
\end{equation}
Substituting (\ref{7}) into (\ref{8}) and solve the linear partial differential equation, we get that
\begin{equation}\label{9}
\xi_i = -\frac{c}{t}\partial_if(x)+\frac{g_i(x)}{t^2},
\end{equation}
where $g_i(x)$ are arbitrary coordinate functions. By using the remaining Killing equations,
we could determine the form of $f(x_i)$ and $g_i(x_i)$, i.e. substituting (\ref{9}) to the following equation
\begin{equation}\label{10}
\partial_i\xi_j+\partial_j\xi_i = 0.
\end{equation}
From these equations we get a general form of the functions $f(x_i)$ and $g_i(x_i)$
\begin{equation}\label{11}
f(x) = a+v_ix^i,  \qquad g_i(x) = \varepsilon_{ijk}x^jp^k+a_i,
\end{equation}
where $a,v_i,p^k,a_i$ are arbitrary constant parameters.
Substituting (\ref{11}) in (\ref{7}) and (\ref{9}) we get
\begin{equation}\label{12}
\xi_0 = \frac{1}{t^2}(a+v_ix_i), \qquad \xi_i = -\frac{c}{t}v_i+\frac{1}{t^2}(\varepsilon_{ijk}x^jp^k+a_i).
\end{equation}
Since
\begin{equation}\label{13}
\xi^0 = g^{00}\xi_0+g^{0i}\xi_i, \qquad \xi^i = g^{i0}\xi_0+g^{ij}\xi_j,
\end{equation}
we get that
\begin{equation}\label{14}
\xi^0 = \frac{c^3t}{R^4}\left(act+a_ix^i\right), \qquad \xi^i = \frac{c^3t}{R^2}\left\{\frac{ax^i}{R^2}+\delta^{ij}v_j\right\}+\frac{c^2}{R^2}\left\{\frac{(a_jx^j)x^i}{R^2}-(a_j+\varepsilon_{jkl}x^kp^l)\right\}.
\end{equation}
Now we can construct the generators of symmetry in the $R$-spacetime using the calculated Killing vectors with superscripts
\begin{equation}\label{15}
X = \xi^\mu\partial_\mu.
\end{equation}
After substituting the equation (\ref{14}) into (\ref{15}) we receive the generators of symmetry in the form
\begin{eqnarray}\label{16}
X = a\frac{c^3}{R^4}t\left\{(t\partial_t+x_i\partial_i)\right\}+\frac{c^3}{R^2}\delta^{ij}v_i(t\partial_j)+ \nonumber \\
+\frac{c^2}{R^2}p^i(\varepsilon_{ijl}\delta^{lk}x^j\partial_k)+\frac{c^2}{R^2}\delta^{ij}a_i\left\{-\partial_j+\frac{\delta_{jk}x^k}{R^2}(t\partial_t+x_l\partial_l)\right\}=
\end{eqnarray}
$$
= \tilde{a}H+\tilde{v^i}K_i+\tilde{a}^iP_i+\tilde{p}^iJ_i,
$$
where $\tilde{a}$,$\tilde{v}_i$,$\tilde{a}^i$,$\tilde{p}^i$ are the transformation parameters which is related with $a$,$v_i$,$a_i$,$p^i$ by
$$
\tilde{a} = \frac{ac^3}{R^4}, \qquad \tilde{v}^i = \frac{\delta^{ij}v_ic^3}{R^2},
$$
$$
\tilde{a}^i = \frac{\delta^{ij}a_jc^2}{R^2}, \qquad \tilde{p}^i = \frac{p^ic^2}{R^2},
$$
and
\begin{equation}\label{17}
H = t(t\partial_t+x^i\partial_i),
\end{equation}
\begin{equation}\label{18}
P_i = -\partial_i+\frac{\delta_{ij}x^j}{R^2}(t\partial_t+x_k\partial_k),
\end{equation}
\begin{equation}\label{19}
K_i = t\partial_i,
\end{equation}
\begin{equation}\label{20}
J_i = \varepsilon_{ijk}x_j\delta^{kl}\partial_l.
\end{equation}

It is important to remark that the algebra of generators (\ref{17})-(\ref{20}) is isomorphic to the algebra of generator of the Poincare symmetry group \cite{02},\cite{07},\cite{11}.
However, only the rotation generators (\ref{20}) are completely the same as in the Poincare symmetry group. The finite coordinate transformations generated by the other generators, are significantly
different from the transformations of Poincare group. The generators (\ref{19}) exactly correspondence to the Galilean boost.
The transformations produced by generators (\ref{17}),(\ref{18}) will be considered in next section.

\section{The symmetry transformations in $R$-spacetime}

We consider the symmetry transformations generated by the generator $P_i$. Let the transformation parameters be $a^i$. We introduce the operator $\hat{Q}$ $:$
\begin{equation*}
\hat{Q} \equiv a^iP_i = -a_i\partial_j + \frac{a_ix^j}{R^2}\left\{t\partial_t + x^k\partial_k\right\}.
\end{equation*}
From now on, $a_i \equiv \delta_{ij}a^j$. Because $\hat{Q}$ is a first-order differential operator, its action on any function $f$
will be the commutator between $\hat{Q}$ and $f$ $:$
\begin{equation}\label{21}
\left(\hat{Q}f\right) = [\hat{Q},f],
\end{equation}
where the parentheses define the action of the operator $\hat{Q}$ on function $f$ only \cite{08}.

Now we consider action of the operator of finite transformation on arbitrary function $f$ $:$
\begin{equation}\label{22}
\tilde{f} \equiv \left(e^{\hat{Q}}f\right),
\end{equation}
where, once again, the parentheses on the right-hand side indicate that the differential
operator $\exp(\hat{Q})$ acts on $f$ only. Since $\left(\hat{Q}f\right) = [\hat{Q},f]$ is
a function, we see that
$\left(\hat{Q}^2f\right) = [\hat{Q},\left(\hat{Q}f\right)] = [\hat{Q},[\hat{Q},f]]$.
More generally, it is easy to see that for arbitrary degree \cite{08}
\begin{equation}\label{23}
\left(\hat{Q}^nf\right) = \underbrace{[\hat{Q},[\hat{Q},[\cdots[\hat{Q},f]\cdots]]]}_n,
\end{equation}
where $n=0,1,2,\ldots$. Thus,
\begin{eqnarray}\label{24}
\left(e^{\hat{Q}}f\right) &=& \sum_{n=0}^\infty \frac{1}{n!}\left(\hat{Q}^nf\right) \nonumber \\
&=& \sum_{n=0}^\infty \frac{1}{n!}\underbrace{[\hat{Q},[\hat{Q},[\cdots[\hat{Q},f]\cdots]]]}_n
\nonumber \\
&=& e^{\hat{Q}}fe^{-\hat{Q}},
\end{eqnarray}
With the help of this formula we are able to find the finite transformation for $t$, $x^i$.
We introduce the variables $\chi \equiv \displaystyle\frac{a_ix^i}{R^2}$ and $\alpha^2 \equiv \displaystyle\frac{a_ia^i}{R^2}$.
The parallel and perpendicular components of $x^i$ to the vector $a^i$ can be correspondingly represented as
\begin{equation}\label{25}
x^i_\| = a^i\frac{\chi}{\alpha^2}
\end{equation}
\begin{equation}\label{26}
x^i_\bot = x^i-x^i_\|
\end{equation}
\begin{enumerate}
\item \underline{Finite transformation of $t$} It is not difficult to calculate the commutators
\begin{equation}\label{27}
[\hat{Q},t] = \chi t
\end{equation}
and
\begin{equation}\label{28}
[\hat{Q},\kappa] = \chi^2-\alpha^2,
\end{equation}
Let $F(x)$ be a differentiable function of $\chi$. We have already noted that the operator $\hat{Q}$ is a first-order differential operator, so we get that
\begin{eqnarray} \label{29}
[\hat{Q},F(\chi)t] &=& \frac{dF}{d\chi}[\hat{Q},\chi]t + F(\chi)[\hat{Q},t] \nonumber \\
&=& \left\{ \left(\chi^2-\alpha^2\right)\frac{dF(\chi)}{d\chi} +\chi F(\chi)\right\}t \nonumber \\
&\equiv & \left(\hat{\cal{D}}F(\chi)\right)t,
\end{eqnarray}
where we have defined the differential operator
\begin{equation}\label{30}
\hat{\cal{D}} \equiv \left(\chi^2-\alpha^2\right)\frac{d}{d\chi}+\chi.
\end{equation}
Since $(\hat{\cal{D}}F)$ is a function of $\chi$, Eq. (\ref{29}) says that the commutator $\hat{Q}$
and some function $F(\chi)$ multiplied by $t$ is again another function of $\chi$ multiplied by $t$.
Since $\left(\hat{\cal{D}}1\right) = \chi$,
this result lead to a simple formula:
\begin{equation}\label{31}
\underbrace{[\hat{Q},[\hat{Q},[\cdots[\hat{Q},t]\cdots]]]}_n = \left(\hat{\cal{D}}^n1\right)t
\end{equation}
Putting (\ref{31}) in the formula (\ref{24}) under the condition $f=t$, we get that
\begin{eqnarray}\label{32}
\left(e^{\hat{Q}}t\right) &=& \sum_{n=0}^\infty \frac{1}{n!}\left(\hat{\cal{D}}^n1\right)t
\nonumber \\
&=& \left(e^{\hat{\cal{D}}}1\right)t.
\end{eqnarray}
To calculate $(\exp(\hat{\cal{D}})1)$, we introduce a new variable $y$ related to $\chi$ by the relation
$$\frac{d\chi}{dy} = \chi^2-\alpha^2$$
or
\begin{equation}\label{33}
\chi = -\alpha\coth(\alpha y)
\end{equation}
Thus, in terms of $y$, the operator $\hat{\cal{D}}$ takes the form
\begin{eqnarray}\label{34}
\hat{\cal{D}} &=& \frac{d}{dy} - \alpha\coth(\alpha y)  \nonumber \\
&=& e^{\int^y \alpha\coth(\alpha z)dz}\, \frac{d}{dy} \, e^{-\int^y \alpha\coth(\alpha z)dz},
\end{eqnarray}
so that the operator of finite transformation come to
\begin{eqnarray}\label{35}
e^{\hat{\cal{D}}} &=& \sum_{n=0}^\infty \frac{1}{n!}\hat{\cal{D}}^n \nonumber \\
&=& e^{\int^y \alpha\coth(\alpha z)dz}\, e^{\frac{d}{dy}} \, e^{-\int^y \alpha\coth(\alpha z)dz}.
\end{eqnarray}
Since
$$  e^{-\int^y \alpha\coth(\alpha z)dz} = \frac{1}{\sinh(\alpha y)},  $$
then
\begin{equation}\label{36}
\left(e^{\hat{\cal{D}}}1\right) = \frac{\sinh(\alpha y)}{\sinh(\alpha(y+1))},
\end{equation}
where we have used the fact that $\exp(d/dy)$ is a translational operator that increases the value
of $y$  by one unit. Using Eq. (\ref{33}), we easily evaluate
\begin{equation}\label{37}
\frac{\sinh(\alpha y)}{\sinh(\alpha(y+1))} = \frac{\sqrt{1-\displaystyle\frac{\rho_i\rho^i}{R^2}}}{1-\displaystyle\frac{\rho_i\rho^i}{R^2}},
\end{equation}
where $\rho^i = \displaystyle\frac{a^i}{\alpha}\tanh\alpha$ (i.e. we note that $\rho_i\rho^i = R^2\tanh\alpha < R^2$).
Finally we obtain the time transformation
\begin{equation}\label{38}
t' = \left(e^{\hat{Q}}t\right) = \frac{t\sqrt{1-\displaystyle\frac{\rho_i\rho^i}{R^2}}}{1-\displaystyle\frac{\rho_i\rho^i}{R^2}}.
\end{equation}
\item \underline{Finite transformation of $x^i_\perp$} Since the commutator
$$ [\hat{Q},x^i_\perp] = \chi x^i_\perp $$
has a form similar to the commutator (\ref{27}), then the finite transformation of $x^i_\perp$ takes the form similar to (\ref{38})
\begin{equation}\label{39}
(x^i_\perp)' = \left(e^{\hat{Q}}x^i_\perp\right) =
\frac{x^i_\perp\sqrt{1-\displaystyle\frac{\rho_i\rho^i}{R^2}}}{1-\displaystyle\frac{\rho_i x^i}{R^2}}.
\end{equation}
\item \underline{Finite transformation of $x^i_\parallel$} Since $x^i_\parallel$ is proportional
to $\chi$, we need to evaluate $(\exp(\hat{Q})\chi)$. Using the commutator (\ref{28}), we calculate
\begin{equation}\label{40}
[\hat{Q},F(\chi)] = \frac{dF}{dy},
\end{equation}
here we use the variable $y$ defined previously in Eq. (\ref{33}). Since the left-hand side of Eq. (\ref{40}) is a function of $\chi$
we get the analogous relation for multiple commutator \cite{10}
\begin{equation}\label{41}
\underbrace{[\hat{Q},[\hat{Q},[\cdots[\hat{Q},F(\chi)]\cdots]]]}_n = \frac{d^nF}{dy^n},
\end{equation}
so that
\begin{equation}\label{42}
\left(e^{\hat{Q}}F(\chi)\right) = \left(e^{\frac{d}{dy}}F(\chi)\right).
\end{equation}
Putting in Eq. (\ref{42}) $F(\chi) = \chi$, we calculate
\begin{equation}\label{43}
\left(e^{\hat{Q}}\chi\right) = \frac{\chi-\alpha\tanh\alpha}{1-\displaystyle\frac{a_ix^i}{\alpha R^2}\tanh\alpha}.
\end{equation}
We finally obtain the finite symmetry transformation of $x^i_\parallel$
\begin{equation}\label{44}
(x^i_\parallel)' = \left(e^{\hat{Q}}x^i_\parallel\right) = \frac{x^i_\parallel-\rho^i}{1-\displaystyle\frac{\rho_i x^i}{R^2}}.
\end{equation}
\end{enumerate}

Now we consider the coordinate transformations created by the generator $H$. We introduce the operator
\begin{equation}\label{45}
\hat{Q} \equiv aH = at(t\partial_t+x^i\partial_i),
\end{equation}
We evaluate the commutator of the generator $\hat{Q}$ with coordinate $x^i$ $:$
\begin{equation}\label{46}
[\hat{Q},x^i] = atx^i,
\end{equation}
and all the following commutators will be
\begin{equation}\label{47}
[\hat{Q},F(t)x^i] = a\left(t^2\frac{dF(t)}{dt}+tF(t)\right)x^i = \left(\hat{G}F(t)\right)x^i,
\end{equation}
where
\begin{equation}\label{48}
\hat{G} \equiv a\left(t^2\frac{d}{dt}+t\right).
\end{equation}
Since $(\hat{G}f(t))$ is a function of $t$ and Eq. (\ref{47}) shows that the commutator $\hat{G}$ with some function $f(t)$ multiplied by $x^i$,
is again a function of $t$ multiplied by $x^i$. Since $\left(\hat{G}1\right) = at$, we obtain the simple relation $:$
\begin{equation}\label{49}
\underbrace{[\hat{Q},[\hat{Q},[\cdots[\hat{Q},x^i]\cdots]]]}_n = \left(\hat{G}^n1\right)x^i,
\end{equation}
Substituting the relation (\ref{49}) to the formula (\ref{24}) with $f = x^i$, we obtain the finite coordinate transformation
\begin{equation}\label{50}
(x^i)' = \left(e^{\hat{Q}} x^i\right) = \sum_{n=0}^\infty\frac{1}{n!}\left(\hat{G}^n 1\right)x^i =
\left(e^{\hat{G}}1\right)x^i.
\end{equation}
To evaluate $\left(e^{\hat{G}}1\right)$ we have to introduce a new variable $z$ connected with $t$ by the relation
\begin{equation}\label{51}
\frac{dt}{dz} = at^2,
\end{equation}
or $t=-1/az$. The operator $\hat{G}$ in new variable takes the form
\begin{equation}\label{52}
\hat{G} = \frac{d}{dz}-\frac{1}{z} = z\frac{d}{dz}z^{-1},
\end{equation}
and we get in the exponential form
\begin{equation}\label{53}
e^{\hat{G}} = \sum_{n=0}^\infty\frac{1}{n!}\left(\hat{G}^n\right) = ze^{\frac{d}{dz}}z^{-1},
\end{equation}
Now we can calculate
\begin{equation}\label{54}
\left(e^{\hat{G}}1\right) = \frac{z}{z+1},
\end{equation}
and finally we obtain the finite transformation
\begin{equation}\label{55}
(x^i)' = \frac{x^i}{1-at}.
\end{equation}
The commutator of generator $\hat{Q}$ with $t$ $:$
\begin{equation}\label{56}
[\hat{Q},t] = at^2
\end{equation}
differs from Eq.(\ref{46}) only by replacing  $x^i$ with $t$. Therefore, all previously carried out calculations can be repeated
for the finite transformation of $t$ and we finally obtain
\begin{equation}\label{57}
t' = \left(e^{\hat{Q}}t\right) = \frac{t}{1-at}.
\end{equation}
Therefore we have obtained all symmetry transformations $:$
\begin{description}
\item[Transformation from $P_i$ $:$]
\begin{equation}\label{58}
t' = \left(e^{\hat{Q}}t\right) = \frac{t\sqrt{1-\displaystyle\frac{\rho_i\rho^i}{R^2}}}{1-\displaystyle\frac{\rho_ix^i}{R^2}},
\end{equation}
\begin{equation}\label{59}
(x^i_\perp)' = \left(e^{\hat{Q}}x^i_\perp\right) = \frac{x^i_\perp\sqrt{1-\displaystyle\frac{\rho_i\rho^i}{R^2}}}{1-\displaystyle\frac{\rho_ix^i}{R^2}},
\end{equation}
\begin{equation}\label{60}
(x^i_\parallel)' = \left(e^{\hat{Q}}x^i_\parallel\right) = \frac{x^i_\parallel-\rho^i}{1-\displaystyle\frac{\rho_ix^i}{R^2}}.
\end{equation}
\item[Transformation from $H$ $:$]
\begin{equation}\label{61}
t'=\frac{t}{1-at},  \qquad (x^i)'=\frac{x^i}{1-at}.
\end{equation}
\item[Transformation from $K_i$ $:$]
\begin{equation}\label{62}
(x^i)' = x^i+v^it,  \qquad t' = t.
\end{equation}
\item[Transformation from $J_i$ $:$]
\begin{equation}\label{63}
(x^i)' = R^i_j x^j,
\end{equation}
\end{description}
where $b^i,a,a^i$ are the transformation parameters and $R^i_j$ is a matrix of rotational group.

\section{Connection with the Lorentz transformations}

At first glance, the previously constructed transformations seem absolutely non-relativistic (i.e. they contain the Galilean transformations in exact form). Moreover the metric of $R$-spacetime (\ref{1}) can be obtained \cite{02} from the metric of anti-de Sitter spacetime in Beltrami coordinates in the non-relativistic limit $c\longrightarrow\infty$.

We show that the group of transformations (\ref{58})-(\ref{63}) indeed contains the transformations which reduces to the ordinary Lorentz transformations in a certain limit \cite{08},\cite{09},\cite{11}. Moreover we note that geodesics in anti-de Sitter spacetime in the Beltrami coordinates are described by linear functions \cite{05}. The group of transformations (\ref{58})-(\ref{63}) are explained by linear-fractional functions which conserve linearity of geodesics. As a result, this group connect inertial reference frames. We will have constructed the connection of spacetime coordinates of two inertial reference frames. We notice that the group of transformations (\ref{58})-(\ref{63}) conserve some invariant subspace -- hypersurface $t=0$. Therefore, it is implied to consider the inertial reference frame where the origin of coordinates coincides in some moment of time, different from zero.

To reduce the number and size of formulas we reduce to considering the longitudinal space transformations. Our purpose is to connect the coordinates of any world point in two different inertial reference frame $(t,x)$
and ($\hat{t},\hat{x}$) correspondingly, with the additional condition that the point $(t=t_0, x=0)$ in coordinates of one inertial reference frame has the coordinates $(\hat{t} = t_0, \hat{x}=0)$ in other reference frame.
For this thing we perform the transformation ($\ref{62}$) with some parameter $v = -\beta R/t_0$ and then  the transformations ($\ref{58}$), (\ref{60}) with $\rho = -\beta/R$.
This will ensure the fulfillment of the condition $\hat{x} = 0$ if $(t=t_0, x=0)$. Actually, under these transformations the arbitrary point $(t,x)$ changes to $(t',x')$ $:$
\begin{equation}\label{64}
x' = \frac{x-\beta R(t-t_0)/t_0}{1+\beta x/R}.
\end{equation}
However, under these transformation the time coordinate changes
\begin{equation}\label{65}
t' = \frac{t\sqrt{1-\beta^2}}{1+\beta x/R},
\end{equation}
and $(t=t_0, x=0)$ change to $(t'=t_0\sqrt{1-\beta^2}, x'=0)$.
To realize the condition $\hat{x} = 0$, we introduce the transformations (\ref{61}) $:$
\begin{eqnarray}
  \hat{t} &=& \frac{t'}{1+at'}, \label{66} \\
  \hat{x} &=& \frac{x'}{1+at'}. \label{67}
\end{eqnarray}
We find a parameter $a$ from the condition $\hat{t}=t_0$ for $t'=t_0\sqrt{1-\beta^2}$. From the relation (\ref{67}) we obtain $a = (1-\gamma)/t_0$,
where $\gamma \equiv 1/\sqrt{1-\beta^2}$. Putting the parameter $a$ and the relations (\ref{64}), (\ref{65}) in right hand side of Eq. (\ref{66}), (\ref{67}),
finally we get
\begin{eqnarray}
  \hat{t}-t_0 &=& \frac{\gamma(t-t_0-\beta t_0x/R)}{1-(\gamma-1)(t-t_0)/t_0+\gamma \beta x/R},  \label{68}\\
  \hat{x} &=& \frac{\gamma(x-\beta R(t-t_0)/t_0)}{1-(\gamma-1)(t-t_0)/t_0+\gamma \beta x/R}.    \label{69}
\end{eqnarray}
So these transformations are little similar to ordinary Lorentz transformations, but exactly they are coincident with the fractional-linear transformations, obtained in works \cite{03},\cite{04}
and called there the Lorentz--Fock transformations.

We show that in small neighborhood of the point $t=t_0, x=0$  the relations (\ref{68}), (\ref{69}) are coincident with Lorentz transformations.
The smallness of  neighborhood implies that $|t-t_0|\ll t_0$ and $|x|\ll R$. The formally-mathematical change to this small neighborhood is realized by
limit change $t_0 \rightarrow \infty, R\rightarrow\infty$ under constant relation $R/t_0\equiv c_0$. Under such limit change the denominator in the relation
(\ref{68}), (\ref{69}) reaches to identity and numerator is brought to the form
\begin{eqnarray}
  \hat{t}-t_0 &=& \gamma(t-t_0-\beta x/c_0),  \label{70}\\
  \hat{x} &=& \gamma(x-\beta c_0(t-t_0)).     \label{71}
\end{eqnarray}
which coincides with ordinary Lorentz transformations. We also note that under $|x|\ll R, |\rho| \ll R$ the transformations (\ref{58}), (\ref{60}) convert to
ordinary space translations.

We consider the transformations (\ref{61}) in small neighborhood of the point $t=t_0$ (i.e. under $|t-t_0|\ll t_0$) and under small value of the transformation parameter $a$
(i.e. under $at_0\ll 1$). It is reached to the limit change $t_0 \rightarrow \infty$ under the finite values $t-t_0 \equiv \tau, at_0^2 \equiv \alpha$.
Under this limit change the relation (\ref{61}) reaches to $\tau' = \tau+\alpha, x'=x$.

Therefore, we have shown that in small neighborhood of origin of space coordinates and away from the hypersurface $t=0$ the constructed transformations coincide
with transformations of Poincare group. Consequently, all physical laws in neighborhood of the observer will coincide in terms of Minkowski spacetime and $R$-spacetime.
However, at cosmological distances and times, significant differences in the interpretation of the observed events are possible \cite{06}.

\end{document}